\documentclass[preprint,showpacs,preprintnumbers,amsmath,amssymb,numerical]{revtex4}

\usepackage{graphicx}
\usepackage{dcolumn}
\usepackage{bm}
\usepackage{color}

\begin{document}


\title{Josephson-like behavior in YBa$_2$Cu$_3$O$_{7-\delta}$ nano-bridges carrying the depairing current}


\author{S. Nawaz$^1$}

\author{R. Arpaia$^{1,2}$}

\author{F. Lombardi$^1$}

\author{T. Bauch$^1$}

\affiliation{$^1$Quantum Device Physics Laboratory, Dept. of Microtechnology and
Nanoscience, Chalmers University of Technology,
SE-412 96 G\"oteborg, Sweden.}

\affiliation{$^2$ CNR-SPIN, Dipartimento di Scienze Fisiche Universit\`{a} degli Studi di Napoli Federico II, Italy}

\date{\today}

\begin{abstract}
We have investigated the zero-field critical supercurrent of YBa$_2$Cu$_3$O$_{7-\delta}$ bridges patterned from 50~nm thick films as a function of bridge width, ranging from 2 $\mu$m to 50~nm. The critical current density monotonically increases for decreasing bridge width even for widths smaller than the Pearl length. This behavior is accounted for by considering current crowding effects at the junction between the bridge and the wider electrodes. Comparison to numerical calculations of the current distributions in our bridge geometries of various widths yields a (local) critical current density at 4.2 K of $1.3\times 10^8$ A/cm$^2$, the Ginzburg Landau depairing current density. The observation of up to 160 Shapiro-like steps in the current voltage characteristics under microwave irradiation substantiates the pristine character of our nano bridges with cross sections as small as $50\times 50$~nm$^2$.
\end{abstract}

\maketitle

\begin{figure}\begin{center}
\includegraphics[width=8.5cm]{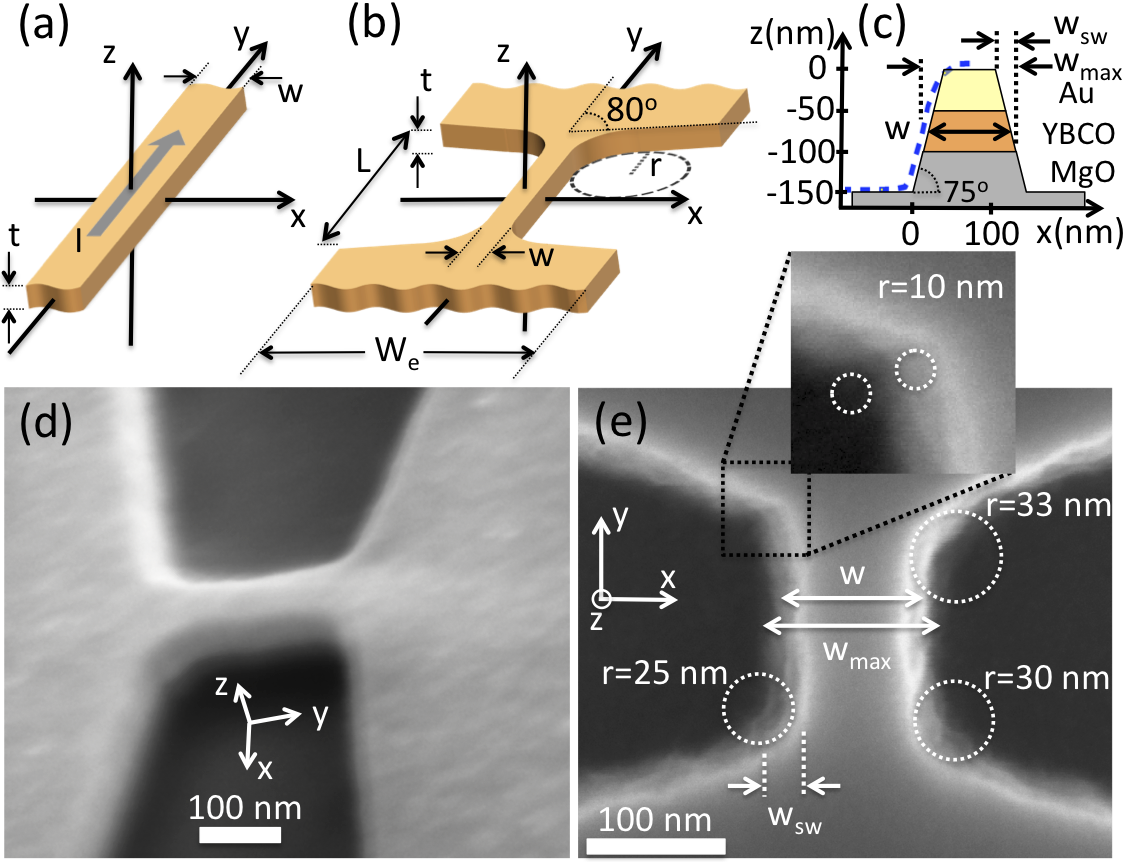}
\caption{(color online){\protect\small \ (a) Sketch of an infinite long bridge of thickness $t$ and width $w$. The current in the bridge flows only along the $y$-direction (b) Sketch of a bridge of thickness $t$, width $w$, and length $L$ connected to wide electrodes.
(c) Sketch of the cross section of a patterned bridge. The dashed line is a typical Atomic Force Microscope (AFM) line scan along the cross section of a nano bridge. The slope of the bridge side walls is $\sim 75^{\circ}$. The lateral extension of the YBCO/Au bridge side walls is given by $w_{sw}\simeq 100$~nm$/ \tan 75^{\circ} = 26$~nm  (d) Scanning Electron Microscopy (SEM) image of a 200~nm long and 75~nm wide bridge (45$^{\circ}$ tilted stage). (e) SEM image of a 200 nm long and 100 nm wide YBCO/Au bridge (top view). The width of the Au/YBCO bridge side wall is $w_{sw}\simeq 25$~nm (see also panel (c)). The dotted circles indicate the bending radii of the inner corners. The inset is a magnification of the upper left inner corner. The two dotted circles denote the bending radii in the Au (right circle) and YBCO film (left circle), respectively.}}
 \label{figure1}
 \end{center}
\end{figure}

Recent advances in nano-patterning techniques have paved the way for studying fundamental aspects of superconductivity on the nanoscale. The expected suppression of superconductivity and the search for quantum coherent phase slip events in superconducting nanowires with cross sectional dimension on the nanometer scale \cite{Bezryadin2008} has triggered a variety of exciting experiments \cite{Mooij2006, Astafiev2012,Arutyunov2012}. The study of nano patterned High critical Temperature Superconductors (HTSs) in the form of nano-rings \cite{Sochnikov2010}, nano-bridges \cite{Mohanty2004, Bonetti2004}, and nano-dots is expected to elucidate the unresolved puzzle of the microscopic mechanism leading to superconductivity in these unconventional materials as recently demonstrated in a nanometer sized HTS island \cite{Gustafsson2013}.  Nanoscale superconductors also allow for new exciting developments towards quantum-limited sensors such as superconducting nanowire single photon detectors (SNSPDs) \cite{Bartolf2010} and nano Superconducting Quantum Interference Devices (nanoSQUIDs) with unprecedented flux sensitivity \cite{Foley2009, Gallop2003}. The realization of wires with highly homogeneous superconducting properties is of essential importance to enable fundamental studies and operational reproducible devices. 
While this issue is within the reach of available nanotechnologies for conventional superconductors \cite{Bezryadin2008}, it still represents a challenge for cuprate HTSs. The chemical instability of these materials, mostly related to oxygen out-diffusion, and the extreme sensitivity to defects and disorder due to the very short superconducting coherence length $\xi$ (of the order of 2~nm), do represent real issues in establishing reliable nanofabrication routines. Indeed, the nano-patterning of HTS materials has been a longstanding challenge.\\
An excellent method for assessing the quality and homogeneity of nano-patterned superconducting bridges is the measurement of the maximum supercurrent density $J_c$ that in bridges with cross sections smaller than the London penetration depth, $\lambda_L$, should be given by the theoretically  expected Ginzburg Landau (GL) depairing limit, $J_{GL}=\Phi_0 / 3\sqrt{3}\pi\mu_0 \lambda_{L}^{2} \xi$, with $\Phi_0\simeq 2\cdot 10^{-15}$Tm$^2$ the superconducting flux quantum, and $\mu_0$ the vacuum permeability. $J_c$ is extremely sensitive to any inhomogeneity in the superconducting properties along the bridge and to the film edge roughness \cite{Tahara1990, Jones2010, Papari2012}. Reaching the theoretical GL depairing limit is an issue even for conventional superconductors \cite{Xu2010}.
Up to now, all experimental values on critical current densities in cuprate HTS nanobridges reported in literature show a wide spread and especially a reduction of the critical current density, $J_c$, when approaching lateral dimensions on the 100~nm scale \cite{Assink1993,Papari2012}. This is indicative of a degradation of the superconducting properties. Moreover, the reported critical current densities are still below the theoretically expected GL depairing limit.\\
In this paper we report an experimental and numerical study of the critical current density, $J_c(w)$, in YBa$_2$Cu$_3$O$_{7-\delta}$ (YBCO) nanobridges as a function of bridge width, $w$, showing that the critical current in our nano bridges is only limited by the GL depairing current density. This limit, never reached earlier for HTS materials, raises also the question about the possibility to establish in such nanostructures a nonlinear supercurrent phase relation so as to detect all the Josephson-like related phenomenology as predicted for superconducting nano bridges with dimensions smaller than the Pearl length \cite{Likharev1972,Likharev1979}.  
We have approached this issue by studying the microwave response of the electronic transport through our nano-bridges. The observation of Shapiro-like steps till the 160$^{th}$ order proves the ac Josephson-like effect in YBCO nano bridges carrying the depairing current.\\
In contrast to previous works, which analyze the experimental data by treating the bridges as infinitesimal long bridges \cite{Tahara1990, Jones2010} (see Fig. \ref{figure1}(a)), here we take into account the influence of the wide electrodes, connecting the nano-bridge to the biasing circuit, on the critical current density (see Fig. \ref{figure1}(b)). Only recently the critical current reduction due to turns and corners in superconducting nanowires  was studied experimentally in conventional superconductors for wire widths much smaller than the Pearl length, $\lambda_P = \lambda_L^2 / t$, where $t$ is the film thickness \cite{Hortensius2012,Henrich2012}. The geometry dependence on the current distribution, e.g. wide electrodes connected to a nano-bridge (see Fig. \ref{figure1}(b)), has been studied theoretically using conformal mapping for structures smaller than the Pearl length \cite{Clem2011}. 
However, since our bridges have widths ranging far below and far above the Pearl length and are connected to electrodes, which are much wider than the Pearl length, we instead apply numerical methods for calculating the current distributions in our structures. 

\begin{figure}\begin{center}
\includegraphics[width=8.6cm]{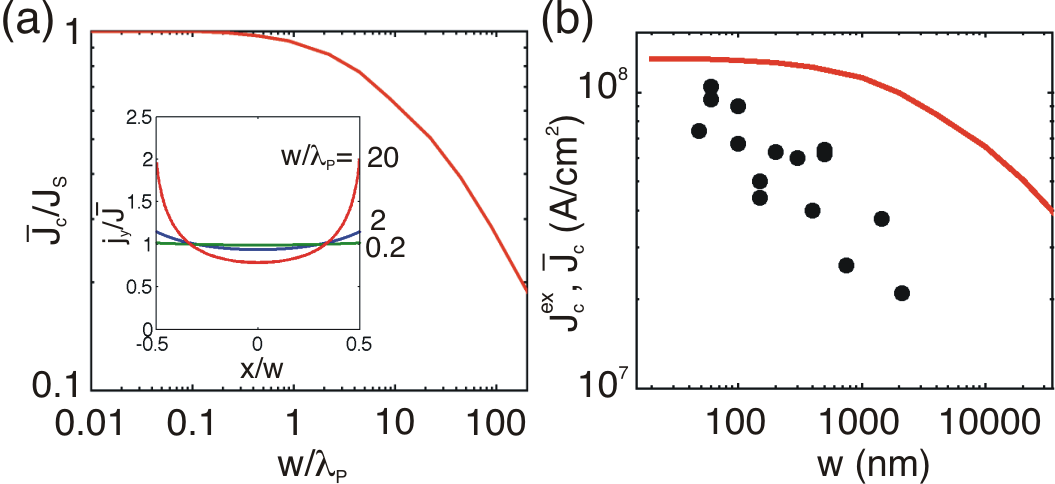}
\caption{(color online){\protect\small \ Calculated normalized average critical current density as a function of bridge width for an infinitesimal long bridge. The inset shows the local current density in $y$-direction across the bridge normalized to the average current density $\overline{J}$ for three different bridge widths: $w=0.2 \lambda_P, 2 \lambda_P$ and $20 \lambda_P$ , respectively. (b) Critical current densities as a function of bridge width, $J_{c}^{ex}(w)$, measured at 4.2 K (symbols). The solid line is the expected critical current density for infinite long bridges using a fixed depairing current density value $J_S=1.3\cdot 10^{8}$ A/cm$^2$. }}
\label{figure2}
 \end{center}
\end{figure}

We fabricated 200 nm long nano-bridges of various widths (50~nm to 2~$\mu$m) from 50 nm thick YBCO films. The YBCO film was grown by Pulsed Laser Deposition (PLD) on a (110) MgO substrate. A 50~nm Au film was deposited ex-situ on top of the YBCO acting as a protective layer for the YBCO film during the patterning process. The patterning of HTS films on the nano scale is an extremely challenging task. The most viable technology is the pattern transfer through a hard mask using Ar ion etching  \cite{Gustafsson2011, Stornaiuolo2010}. However, the detrimental effect of the Ar ion etching on the exposed surfaces of YBCO causes damaged layers having reduced superconducting or even insulating properties \cite{Herbstritt2002}. We drastically improved the nano-patterning of YBCO obtaining nano bridges without any deterioration of the superconducting properties, as we will show below. This has been achieved by using electron beam lithography in combination with a 100 nm thick carbon mask and a very ÒgentleÓ ion milling to define the nanobridges \cite{Gustafsson2011, Nawaz2012}. Here we used an ion acceleration voltage close to the threshold value of  $V\simeq 300$~V, below which YBCO is not etched. Moreover we have used the lowest ion beam current density $J_{Ar^{+}} =0.08$~mA/cm$^2$ that allowed the ignition of the plasma in our milling system. 
The total etching time is such that we also ion-mill approximately 50~nm into the substrate. This assures the removal of any redeposited YBCO in the vicinity of the nano bridge. Scanning Electron Microscopy (SEM) images of typical nano bridges are shown in Fig. \ref{figure1}(d) and (e). An Atomic Force Microscope (AFM) line-scan along the cross section of a typical nano bridge together with a sketch of a bridge cross section is shown in Fig. \ref{figure1}(c). We define the width of a bridge, $w$, as the bridge width at half the YBCO film thickness (see Fig. \ref{figure1}(c)). The width of the YBCO/Au bridge side walls determined from AFM, $w_{sw}\simeq 26$~nm (see Fig. \ref{figure1}(c)), is in good agreement with the one determined from SEM (Fig. \ref{figure1}(e)). Thus, we can determine the width of a bridge from SEM, $w = w_{max}-w_{sd}$, where $w_{max}$  is the width of the bridge at the interface between YBCO and MgO (see Fig. \ref{figure1}(e)) \cite{note1}.
The electrical transport measurements of our nano-bridges were performed in a $^3$He cryostat, which is placed in an electromagnetically shielded room. The current voltage characteristics (IVC) were recorded using a 4-point measurement scheme. While ramping the bias current we simultaneously monitor the current through the bridge and the voltage drop across the bridge. 
All nano-bridges exhibit critical temperatures similar to that of the wide electrodes, $T_c  \simeq 85$~K, differing not more than 1~K (data not shown).
The critical current of the bridge $I_{c}^{ex}$ is determined from the IVC as the bias current above which the bridge undergoes a transition from the zero voltage state to the finite voltage state. Here we use a voltage criterion of 2 $\mu$V. From the critical current values we can calculate the average critical current density $J_{c}^{ex} = I_{c}^{ex}/A_{cr}$ for each bridge, where $A_{cr} = w \times t$ is the smallest cross sectional area of the bridge, which we determined by SEM imaging . Fig. \ref{figure2}(b) shows the experimentally determined $J_{c}^{ex}$ as a function of bridge width. One can clearly observe a monotonic increase of $J_{c}^{ex}$ for decreasing bridge width $w$.\\ 
At first we discuss the width dependence of the critical current density, $J_c(w)$, in infinite long (type II) superconducting bridges, i.e. neglecting the influence of wide electrodes (see Fig. \ref{figure1}(a)), and compare it to the experimental $J_c^{ex}(w)$ dependence measured on our YBCO nanobridges. We limit ourselves to the thin film case ($t<\lambda_L$) since the thickness of our c-axis films ($t=50$~nm) is well below the in plane London penetration depth $\lambda_{L}^{ab}\simeq 150-220$~nm \cite{Papari2012}. Thus we can neglect current components parallel to $z$-direction and assume a homogeneous current distribution throughout the whole film thickness. In addition, all the lateral dimensions of our bridges are larger than the superconducting coherence length $\xi\simeq 1.5-2$~nm. In this limit vortex dynamics across the bridge determines the critical supercurrent. 
We consider the case of zero externally applied magnetic field and account only for the self-field caused by the transport current. For bias currents smaller than the critical current, $I_c$, a finite edge barrier prevents vortices from entering the bridge \cite{Bean1964, Kupriyanov1975,  Aslamazov1983, Benkraouda1998, Maksimova1998, Vodolazov1999, Elistratov2002, Gurevich2008, Bulaevskii2011}. By increasing the bias current from zero to a finite value gradually reduces this edge barrier.  For bias current approaching the critical current   the barrier is eventually completely suppressed at a distance on the order of the coherence length from the bridge edge, allowing vortices to enter the bridge. The resulting vortex motion across the bridge, driven by the Lorentz force, causes a finite voltage drop along the bridge.
The value of the critical supercurrent depends on the detailed in plane current distribution $\vec{j}(x,y)$ in the bridge, which for $\xi\ll\lambda_L$ is given by the Maxwell and London equations \cite{Khapaev1997}: 
\begin{equation}\label{eqn1}
    \mu_0\vec{\nabla}\times ( \lambda^2\cdot\vec{j} ) + \vec{B} =0; \;\;
    \vec{\nabla}\times \vec{B} = \mu_0 \vec{j},
\end{equation}
where $\mu_0$ is the vacuum permeability, $\lambda^2$ the material specific London penetration depth (squared) tensor, and $\vec{B}$ the magnetic field solely generated by the transport currents. The equation set \ref{eqn1} describes the Meissner state, i.e. no static Abrikosov vortices are present in the film.
For bridges having a width smaller than the Pearl length, $\lambda_P$, the current distributes homogeneously across the width. The resulting (average) critical current density value $J_c = I_c/wt$ corresponds approximately to the one-dimensional GL depairing current density, $J_{GL}$ \cite{Bulaevskii2011}.\\
For bridge widths much larger than the Pearl length the current density in $y$-direction is inhomogeneous across the width of the bridge, $j_y(x) \neq const$. This is a direct consequence of the Meissner state keeping the center of the bridge field free.\\
\begin{figure}\begin{center}
\includegraphics[width=6cm]{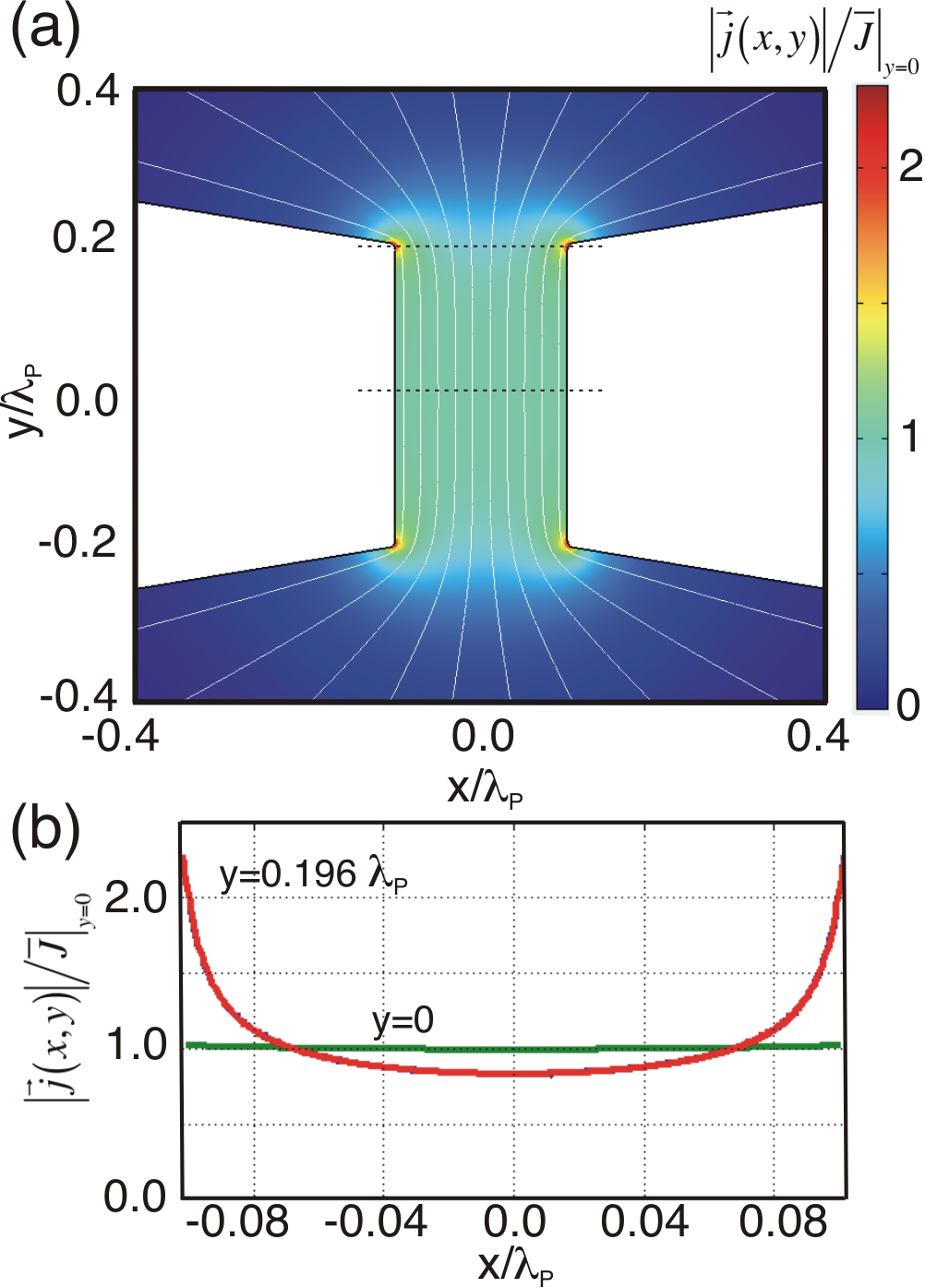}
\caption{(color online){\protect\small \ (a) Calculated absolute value of the local current density normalized to the average current density at $y=0$. The width of the bridge is $w=0.2\lambda_P$ and the length $L=0.4\lambda_P$. The inner corners have a bending radius $r=0.05w$. The width and the length of the electrodes are $10\lambda_P$ and $12\lambda_P$, respectively. A constant current density is injected at the end of one electrode at $y=12.2\lambda_P$ and extracted at the other electrode at $y=-12.2\lambda_P$ (not shown). The white lines indicate the path of the current flow. (b) Line cuts of the current density at $y=0$ and $y=0.196\lambda_P$ indicated as dashed lines in (a). }}
 \label{figure3}
 \end{center}
\end{figure}
We calculated the current distributions $j_y(x)$ for various bridge widths by numerically solving equation set \ref{eqn1} on a geometry depicted in Fig.\ref{figure1}(a). In the inset of Fig. \ref{figure2}(a) we show $j_y(x)$ for three different bridge widths, $w=0.2 \lambda_P, 2 \lambda_P$  and  $20 \lambda_P$, respectively. One can clearly observe that the current density at the edges of a bridge with $w>\lambda_P$ is enhanced compared to the average value $\overline{J}=\int j_y(x)dx/w$. When ramping up the bias current applied to a bridge the critical supercurrent is reached once the local current density at the edges of the bridge $j_y(\pm w/2)$ equals a value close to the depairing current density, $J_{GL}$ \cite{Aslamazov1983, Vodolazov1999}. At this point the edge barrier is suppressed and vortices can enter the bridge, causing a transition from the zero voltage state to the finite voltage state. Thus, from the local current density at the edges $j_y(\pm w/2)$ and the average current density $\overline{J}$ one can compute the average critical current density as a function of width:  
\begin{equation}\label{eqn2}
    \overline{J}_c(w) = J_S\cdot\overline{J}/j_y(\pm w/2),
\end{equation}
where $J_S$ is approximately the depairing value, $J_S\simeq J_{GL}$. In Fig. \ref{figure2}(a) we show the resulting normalized average critical current density $\overline{J}_c(w)/J_S$ as a function of bridge width.\\

\begin{figure}\begin{center}
\includegraphics[width=6.5cm]{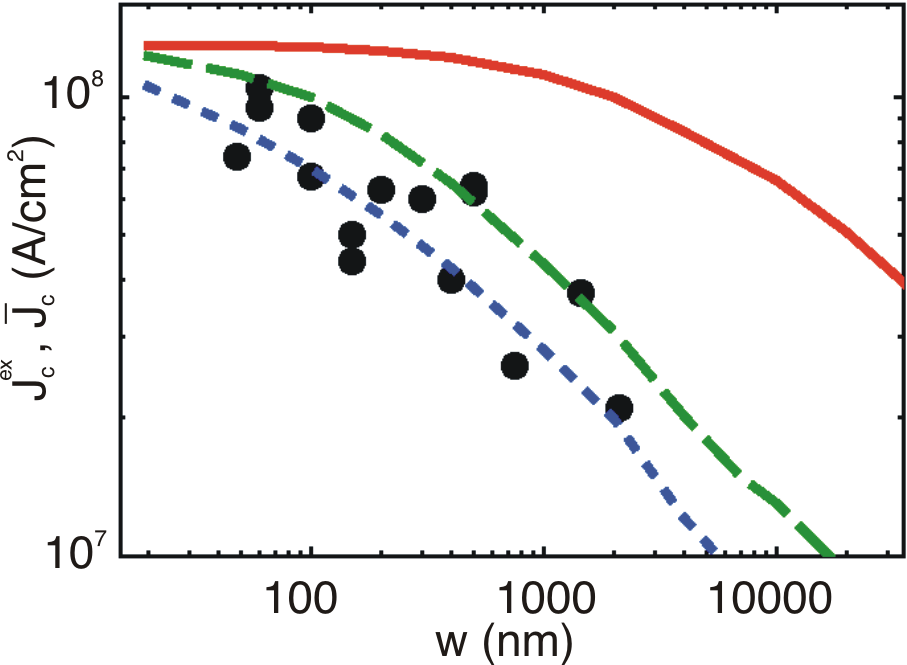}
\caption{(color online){\protect\small \ Critical current density as a function of bridge width measured at 4.2 K (symbols). The solid line is the numerically calculated critical current density for an infinite long bridge using equation \ref{eqn2}. The dashed and dotted lines are the numerically calculated critical current densities using equation \ref{eqn3} for 200~nm long bridges connected to wide electrodes with inner corner radii $r_1=40$~nm and $r_2=10$~nm, respectively.}}
 \label{figure4}
 \end{center}
\end{figure}
In Fig. \ref{figure2}(b) we show the expected $\overline{J}_c(w)$ for infinite long bridges (solid line) together with the experimentally determined critical current densities $J_c^{ex}$ (symbols) as a function of bridge width.
One can clearly see that $J_{c}^{ex}$  does not saturate for bridge width below the Pearl length $\lambda_P\simeq 800$~nm.
The discrepancy between the experimental data and the expected behavior for infinite long bridges, however, can be explained by taking into account the influence of the on-chip (wide) electrodes connecting the bridge to the bias circuitry (see Fig. \ref{figure1}(b)). In this case the current injection from the wide electrodes into the thin bridge causes current crowding at the inner corners of the junction between bridge and electrode, i.e. the local current density at the inner corners is enhanced compared to the average current density at the center of the bridge \cite{Hagedorn1963,Clem2011}, $\overline{J}|_{y=0} = \int j_y(x,y=0)dx/w$. To illustrate the current crowding we show in Fig. \ref{figure3}(a) a calculated local current density in a typical geometry depicted in Fig. \ref{figure1}(b) by solving numerically the equation set \ref{eqn1}. Fig. \ref{figure3}(b) shows two line cuts of Fig. \ref{figure3}(a): one at the center of the bridge and the other close to the electrodes.  One can clearly see the enhanced current density at the inner corners of the bridge/electrode geometry even though the width of the bridge is much smaller than the Pearl length, $w = 0.2\lambda_P$.
The average critical current density in this case can be computed from the numerically determined current distributions in the following way:
\begin{equation}\label{eqn3}
    \overline{J}_c(w) = J_S\cdot\overline{J}|_{y=0}/j_{max},
\end{equation}
where $j_{max}$ is the maximum value of the current density located at the inner corners. 
Depending on the ratio between the inner corner bending radius and the bridge width, the current crowding can strongly reduce the average critical current density of a bridge below its infinite long bridge limit even for bridge widths smaller than the Pearl length.\\
\begin{figure}\begin{center}
\includegraphics[width=8.5cm]{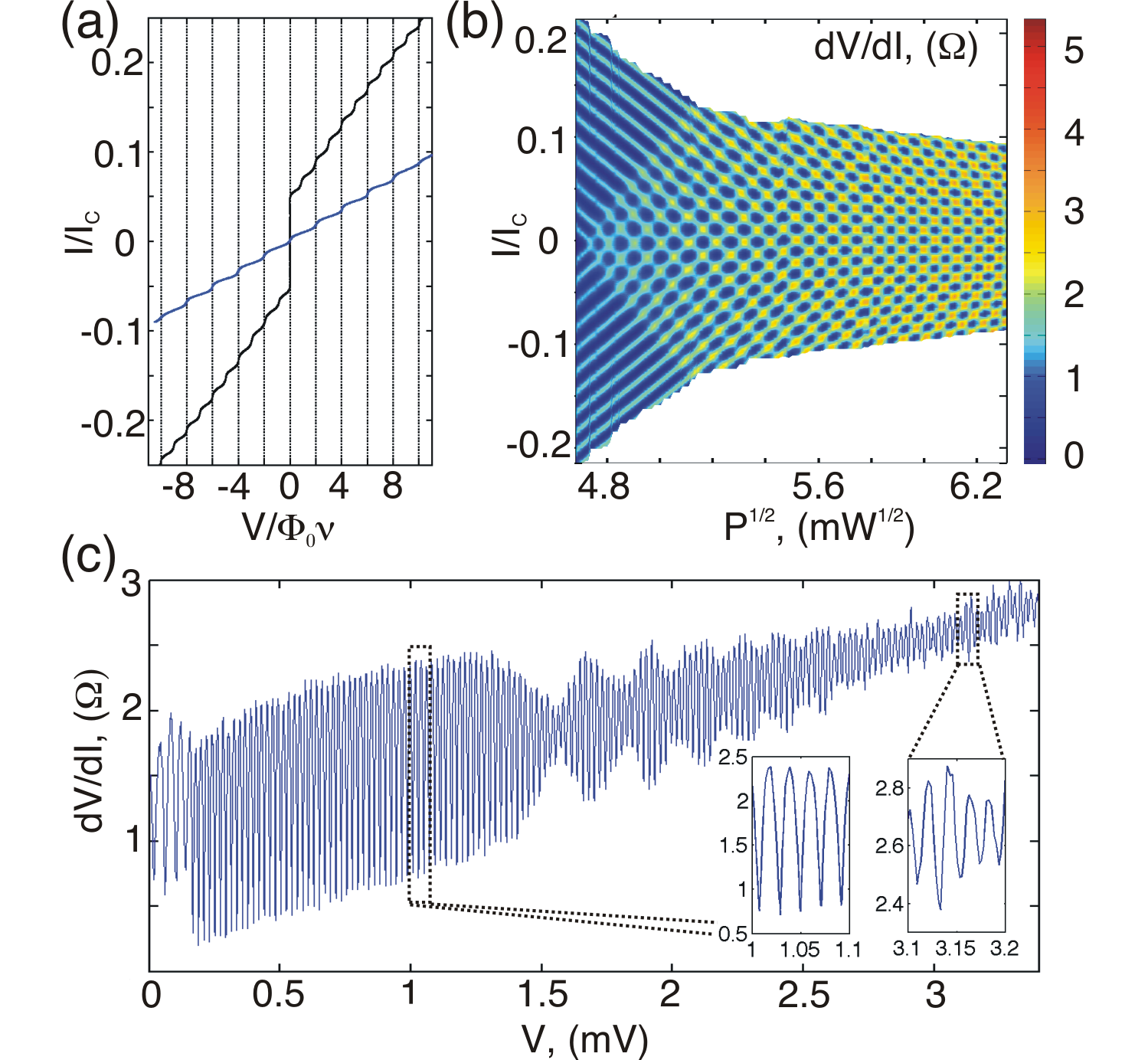}
\caption{(color online){\protect\small \  Shapiro-like steps. (a) Current voltage characteristic at two applied microwave powers (source output), $P=13$dBm (black line) and $P=$16dBm (blue line). (b) Differential resistance map as a function of applied microwave amplitude (at source output) and bias current. The dark blue regions correspond to the Shapiro-like (current) steps. (c) Differential resistance as a function of voltage at a fixed microwave power $P=14.5$dBm. The insets show close-ups of the Shapiro-like steps. In all three panels the temperature is T = 4.2 K and the applied microwave frequency $\nu=10.13$GHz.}}
 \label{figure5}
 \end{center}
\end{figure}
In Fig. \ref{figure4} we show the experimental $J_{c}^{ex}$ dependence (symbols) together with the numerically calculated average critical current densities of bridges having various widths for two different bending radii of the inner corners, $r$, using equation \ref{eqn3}. Here we chose $r_1=40$~nm (dashed line) and $r_2=10$~nm (dotted line). These values comprise the range of typical bending radii we obtain by our lithography process for nominally $80^{\circ}$ corners (see Fig. \ref{figure1}(e)). For comparison we also plotted the numerically calculated $\overline{J}_c(w)$ for the infinite long bridge case (solid line, see also Fig. \ref{figure2}). For all the numerically determined $J_c(w)$ dependencies we used fixed values for the Pearl length, $\lambda_P=800$~nm, and $J_S=1.3\cdot 10^8$A/cm$^2$. The agreement between experimental data and the current crowding model is indeed very good. Moreover the value of the local critical current density $J_S$ used to fit our measurements is very close to the maximum theoretical depairing current density for YBCO. This reflects the excellent quality of our nanobridges with cross sections as small as  $50 \times 50$~nm$^2$. It is worth noting that the critical current density of a nanobridge connected to wide electrodes is approaching the depairing value only for width smaller than half the bending radius of the inner corners (see dashed line in Fig. \ref{figure4}). A similar result was obtained for geometries much smaller than the Pearl length using a conformal mapping approach \cite{Clem2011}.

Having shown that the critical current of our YBCO nanobridges are indeed only limited by the GL depairing current density we furthermore studied the ac Josephson-like effect in our smallest bridges with cross section $50\times50$~nm$^2$. If a superconducting nanobridge having dimensions smaller than the Pearl length is exposed to a microwave field, current steps in the IVC may appear at specific voltage values $V_n = n\cdot \nu \cdot \Phi_0$, where $n$ is an integer, $\nu$ is the applied microwave frequency, and $V$ is the voltage drop along the bridge \cite{Likharev1972, Aslamazov1975}. These steps appear due to the synchronization of the coherent motion of Abrikosov vortices to the microwave radiation frequency by phase locking.  Such current steps are similar to Shapiro steps observable in the IVC of Josephson tunnel junctions when an external microwave field phase locks with the Josephson oscillations at finite voltages \cite{Shapiro1963}. In Fig. \ref{figure5}(a) we show the measured IVC under microwave irradiation ($\nu = 10.13$GHz) for two different applied microwave powers. The Shapiro-like (current) steps occur at integer multiples of  $\nu \Phi_0$. A differential resistance map of the nanobridge as a function of bias current and applied microwave amplitude is shown in Fig. \ref{figure5}(b), where the Shapiro-like steps appear as dark blue regions. The periodic modulation of the current steps with the applied microwave amplitude gives a strong indication of the existence of an effective periodic current phase relation in the nanobridges \cite{Bae2009}. 
The observation of up to 160 Shapiro steps, shown in Fig \ref{figure5}(c), further corroborates the excellent quality of our bridges since any inhomogeneity in the superconducting properties would inhibit the coherent motion of Abrikosov vortices and consequently the Josephson-like behavior \cite{Likharev1972}. The maximum voltage $V_{max}\simeq 3$mV at which we can detect Shapiro-like steps correspond to a Josephson-like oscillation frequency of 1.5 THz making these bridges interesting for applications in detecting and mixing of THz signals and possibly as THz radiation sources.\\ 
In conclusion we have performed a systematic study of the critical current of YBCO bridges, patterned from a 50~nm thick film, as a function of bridge width ranging from 2~$\mu$m to 50~nm. All our bridges can be characterized by a (local) critical current density approaching the Ginzburg Landau depairing critical current density, $1.3\cdot 10^8$A/cm$^2$, down to cross sections of $50 \times 50$~nm$^2$.
The excellent quality of our YBCO nano bridges opens the way for foundational studies of nano-patterned unconventional superconductors where nanoscale ordering has a crucial role in building up the superconducting ground state \cite{Mohanty2004, Bonetti2004}.
Moreover, the pristine superconducting properties of our bridges enable the realization of nanosized quantum limited detectors such as YBCO nanoSQUIDs operational in a wide temperature range. The observed current crowding effect in our nano bridges has furthermore strong implications for the design of superconducting nanobridge based single photon detectors since a homogeneous current density along the whole bridge is essential for improving photon detection sensitivity \cite{Bartolf2010}. The Josephson-like behavior of our nanobridges makes them also attractive for applications in the THz regime.\\
This work has been partially supported by the Swedish Research Council (VR) and the Knut and Alice Wallenberg Foundation.


\begin{thebibliography}{40}
\bibitem[{not()}]{note1}
\bibinfo{note}{From the normal resistance of bare YBCO nanobridges (after
  removal of the gold capping layer) we obtain a sheet resistance of 20
  {$\Omega$} slightly above Tc corresponding to a resistivity {$\rho\simeq$ 100
  $\mu\Omega$cm}, which is a typical value for thin YBCO films.}
\end{thebibliography}

\end{document}